\documentclass[journal]{IEEEtran}
\usepackage{cite}
\usepackage{amsmath,amssymb,amsfonts}
\usepackage{algorithmic}
\usepackage{graphicx}
\usepackage{textcomp}
\usepackage{balance}
\usepackage{multirow}

\begin{document}
\title{ESI-GAL: EEG Source Imaging-based Kinematics Parameter Estimation for Grasp and Lift Task}
\author{Anant Jain and Lalan Kumar
\thanks{Anant Jain is with the Department of Electrical Engineering, Indian Institute of Technology Delhi, New Delhi 110016, India (e-mail: anantjain@ee.iitd.ac.in).}
\thanks{Lalan Kumar is with the Department of Electrical Engineering, Bharti School of Telecommunication, and Yardi School of Artificial Intelligence, Indian Institute of Technology Delhi, New Delhi 110016, India (e-mail: lkumar@ee.iitd.ac.in).}}

\maketitle

\begin{abstract} Electroencephalogram (EEG) signals-based motor kinematics prediction (MKP) has been an active area of research to develop brain-computer interface (BCI) systems such as exosuits, prostheses, and rehabilitation devices. However, EEG source imaging (ESI) based kinematics prediction is sparsely explored in the literature. In this study, pre-movement EEG features are utilized to predict three-dimensional (3D) hand kinematics for the grasp-and-lift motor task. A public dataset, WAY-EEG-GAL, is utilized for MKP analysis. In particular, sensor-domain (EEG data) and source-domain (ESI data) based features from the frontoparietal region are explored for MKP. Deep learning-based models are explored to achieve efficient kinematics decoding. Various time-lagged and window sizes are analyzed for hand kinematics prediction. Subsequently, intra-subject and inter-subject MKP analysis is performed to investigate the subject-specific and subject-independent motor-learning capabilities of the neural decoders. The Pearson correlation coefficient (PCC) is used as the performance metric for kinematics trajectory decoding. The rEEGNet neural decoder achieved the best performance with sensor-domain and source-domain features with a time lag and window size of $100$ $ms$ and $450$ $ms$, respectively. The highest mean PCC values of 0.790, 0.795, and 0.637 are achieved using sensor-domain features, while 0.769, 0.777, and 0.647 are achieved using source-domain features in x, y, and z-directions, respectively. This study explores the feasibility of trajectory prediction using EEG sensor-domain and source-domain EEG features for the grasp-and-lift task. Furthermore, inter-subject trajectory estimation is performed using the proposed deep learning decoder with EEG source domain features.
\end{abstract}


\section{Introduction}\label{sec:introduction}

\subsection{Background}\label{sec:motivation}

Brain$-$computer interface (BCI), also known as brain$-$machine interfaces (BMIs), represent a cutting-edge technology that facilitates control over external devices through brain activity, bypassing the need for peripheral nerves and muscles \cite{wolpaw2020brain}. This technology holds significant promise for enhancing the life quality of individuals with motor disabilities \cite{mane2020bci,miladinovic2020evaluation, na2021embedded, cao2022effective, catalan2023hybrid, demarest2024novel} and enables interaction with healthy individuals \cite{hekmatmanesh2021review,
nijholt2022brain, moreno2023assessing}. Over time, advancements in computational power and machine learning algorithms have empowered BCI systems to decode neural signals for assisting, augmenting, or restoring motor functionality \cite{robinson2021emerging}. BCI systems encompass a series of sequential procedures comprising the acquisition and processing of neural signals, relevant feature extraction, intention detection, and user feedback signals generation. Invasive and non-invasive recording techniques are utilized for capturing neural activity. While invasive approaches offer greater accuracy in brain activity recognition, they necessitate surgical implantation of sensors beneath the scalp \cite{millan2010invasive, miller2020current}. Conversely, non-invasive BCIs capture neural signals by pacing sensors on the scalp. Various non-invasive techniques include functional near-infrared spectroscopy (fNIRS) \cite{naseer2015fnirs}, magnetoencephalography (MEG) \cite{wang2024unilateral}, functional magnetic resonance imaging (fMRI) \cite{baqapuri2021novel} and electroencephalogram (EEG) \cite{gu2021eeg}.

EEG-based BCI systems have gained popularity among these methods due to their high temporal resolution, portability, and cost-effectiveness. They have been utilized across various applications, such as fatigue and drowsiness detection \cite{zheng2022new, othmani2023eeg}, emotion recognition \cite{li2022eeg, jafari2023emotion}, wearable exoskeletons \cite{tang2023upper, li2024human}, and robotic control \cite{cheng2022robotic, ai2023bci, zhou2023shared, fu2024control}. Additionally, EEG-based BCIs have been employed for classifying motor imagery or motor execution tasks \cite{gu2020eeg, altaheri2023deep, wang2023eeg, liu2023multiwavelet}. Although classification-based approaches have been extensively explored, continuous kinematic estimation-based approaches offer enhanced performance and efficient control of external devices such as neural prostheses, exosuits, or exoskeletons. 

\subsection{Related work}\label{sec:literature}
EEG-based kinematics decoding has been performed in both motor execution task \cite{bradberry2010reconstructing,presacco2011neural,ofner2012decoding} and motor imagery task \cite{korik2018decoding}. EEG-based kinematics decoding is performed for 2D center-out reaching task in \cite{sun2017} using multiple linear regression (mLR) decoder. EEG slow cortical potentials based upper limb trajectories (hand, elbow, and shoulder) decoding is reported in \cite{sosnik2020reconstruction} using mLR decoding model. Hand kinematics decoding for unimanual target-reaching movement is investigated in \cite{robinson2021} with Kalman filter (KF) and mLR decoders using scalp EEG signals. Recent studies have utilized deep learning-based decoders \cite{shakibaee2019decoding, jeong2020brain, jain2k22premovnet, pancholi2k21, jain2022subject, saini2023bicurnet} for motor kinematics prediction (MKP) using scalp EEG signals. Knee joint angle trajectory is estimated using EEG signals with the NARX neural network in \cite{shakibaee2019decoding}. EEG-based MKP is investigated in \cite{jeong2020brain} using a convolutional neural network - bidirectional long short-term memory (CNN-biLSTM) decoding model during unimanual target-reaching movement for controlling a robotic arm. MKP is performed for the grasp-and-lift task using a wavelet packet decomposition (WPD) based CNN-LSTM decoder in \cite{pancholi2k21}. Low-frequency EEG signals are utilized in \cite{jain2k22premovnet} for MKP with CNN-LSTM decoding architecture. The inter-subject decoding analysis is investigated in \cite{jain2022subject} using EEG signals along with deep learning-based decoding models. Elbow joint angle trajectory decoding is performed in \cite{saini2023bicurnet} using an attention-based CNN-LSTM decoder with EEG signals during the biceps-curl task.

In the field of brain-computer interface (BCI) research, non-invasive investigations have traditionally focused on utilizing EEG signals directly, operating within the sensor space. However, advancements in EEG source imaging (ESI) techniques enable the inference of cortical sources from non-invasive brain signals. This approach has prompted studies exploring decoding brain activity in the source space via ESI. The application of source-space decoding has predominantly been investigated within the classification domain, encompassing tasks such as motor imagery classification \cite{edelman2015eeg, li2019decoding, hou2020novel}, arm direction classification \cite{handiru2017eeg} and gesture (alphabets) recognition \cite{neuroair2024}. Across these studies, source-space classification has frequently demonstrated enhanced performance compared to traditional sensor space methods. ESI-based MKP has been explored for reach-to-target task \cite{sosnik2021reconstruction}, snake trajectory tracking \cite{srisrisawang2022applying} and grasp-and-lift task \cite{jainembc2023}. In \cite{sosnik2021reconstruction}, the mLR decoder is deployed to estimate hand, elbow, and shoulder trajectories during a target-reaching experiment utilizing EEG current source dipoles. The study reported an average correlation of 0.36 $\pm$ 0.13 across all trajectories for actual movement execution. Source-space EEG-based hand trajectory estimation is conducted by employing a combination of the partial least squares (PLS) regression model and the square-root unscented Kalman filter (SR-UKF) in \cite{srisrisawang2022applying}. The study reported the highest average correlation of 0.31 ± 0.09 and 0.35 ± 0.09 for hand position and velocity, respectively. In \cite{jainembc2023}, hand trajectory estimation is performed for the grasp-and-lift task using EEG source-domain signals with a residual CNN-LSTM neural decoder. The mean Pearson correlation coefficient of 0.59, 0.61, and 0.56 is reported for the window size of 300 ms in the x, y, and z directions, respectively. 


\begin{figure*}[t]
	\centering
	\includegraphics[width=0.92\textwidth]{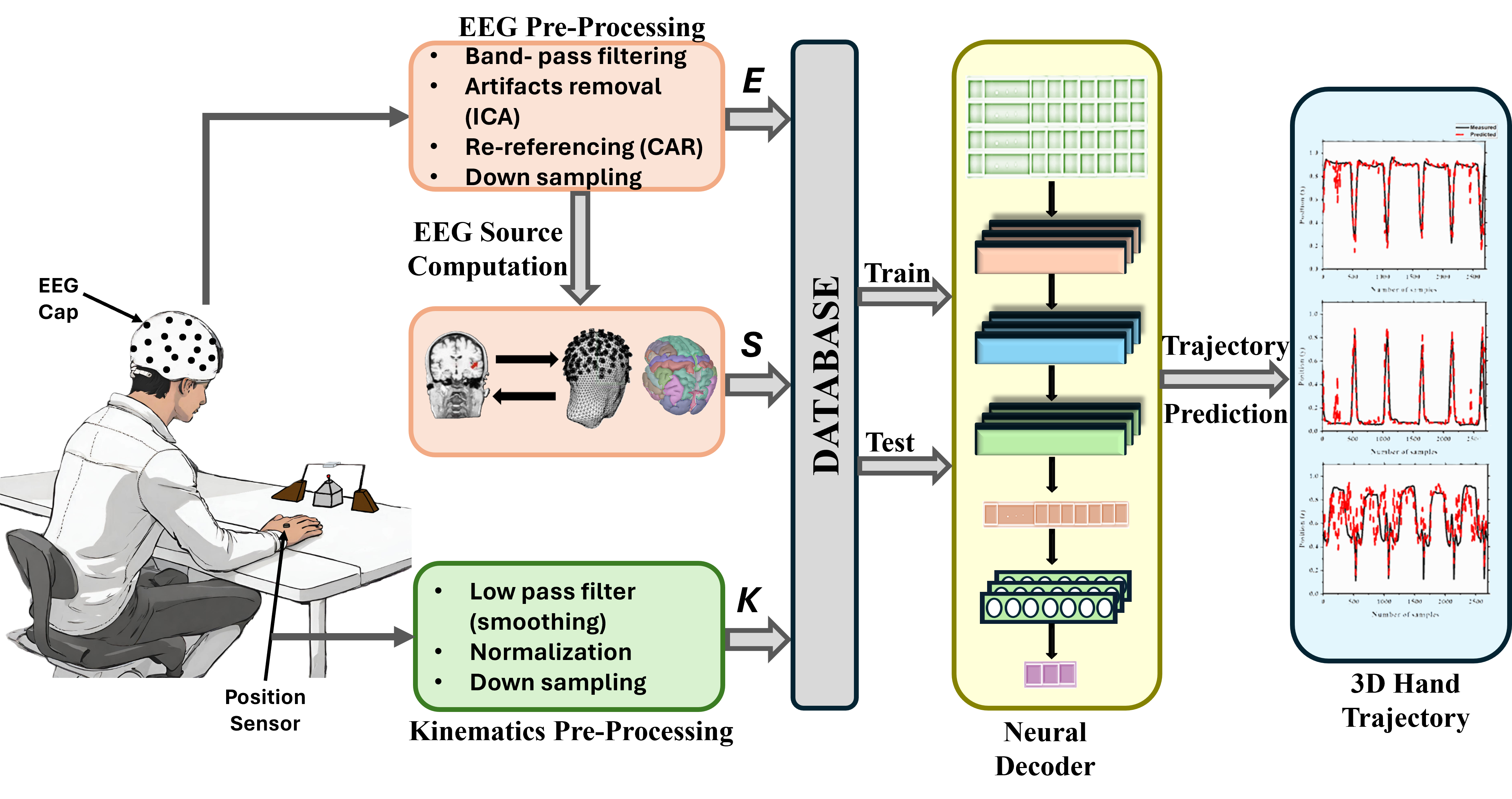}
        \centering
	\caption{Flowchart of proposed kinematics decoding framework for Grasp-and-Lift task.}
	\label{blockdia} 
\end{figure*}
\subsection{Objectives and Contribution}\label{sec:obj&contri}
In this study, MKP is performed using sensor-domain and source-domain EEG features for the grasp-and-lift task. Premovement EEG segments are utilized for kinematics estimation along with convolutional neural network (CNN) based deep learning models for external device (exosuit, prosthesis) control-based BCI applications. The open-source WAY-EEG-GAL \cite{luciw2014multi} database is used for the proposed methodology. The sensor-domain and source-domain features are explored for motor kinematics prediction during the grasp-and-lift task. The frontoparietal regions are utilized as regions of interest (ROI) for both sensor-domain and source-domain-based feature selection. Additionally, inter-subject decoding performance is evaluated to demonstrate the subject-adaptability of the proposed kinematics decoding framework.

The organization of the article is as follows: Section \ref{sec:m&m} includes the dataset description (Section \ref{sec:exppara}), data preprocessing (Section \ref{sec:prepro}), EEG source Imaging and feature extraction (Section \ref{sec:esidata}), deep-learning models (Section \ref{sec:decoders}), and experimental details (Section \ref{sec:expdetails}). Performance evaluation for MKP and an extensive discussion of the results are detailed in Section \ref{sec:r&d} and Section \ref{sec:conclusion} provides the conclusions about the research work.
\section{Materials and Methods}\label{sec:m&m}

This Section includes a detailed description of the experimental paradigm, dataset, data pre-processing pipeline, feature extraction, EEG source imagining, and neural decoders for trajectory estimation. An overview of the proposed trajectory estimation framework is shown in Fig. \ref{blockdia}.

\subsection{Dataset Description}\label{sec:exppara}
WAY-EEG-GAL (Wearable interfaces for
hAnd function recovery EEG grasp and lift) dataset\cite{luciw2014multi} is used for MKP. Synchronous EEG and hand kinematics data were collected during the grasp-and-lift task. EEG data was recorded using 32-channel active EEG electrodes (actiCAP, \textit{Brain Products GmbH, Germany}) in conjunction with a biosignal amplifier (BrainAmp, \textit{Brain Products GmbH, Germany}) at a 500 Hz sampling rate. The hand kinematics data was acquired using a 3D position sensor (FASTRAK, \textit{Polhemus Inc., USA}) with a 500 Hz sampling frequency. In particular, the neural and kinematics data of twelve participants while performing the right-hand grasp-and-lift task is included in the dataset.

In each trial, participants were instructed to perform a grasp-and-lift task on an object. The task involved reaching for a small object, gripping it between the index finger and thumb, elevating it a few centimeters into the air, maintaining a stable hold briefly, and replacing and releasing it to its original position. The initiation of the reaching motion and the controlled lowering of the object were prompted by a visual cue in the form of an LED signal. The participants had autonomy over the overall pace of the task, allowing for a naturalistic execution. Throughout the experimental sessions, the object's properties underwent systematic variations in physical properties. These variations encompassed unpredictable changes in weight (165, 330, or 660 g) and the contact surface (sandpaper, suede, or silk).

\subsection{Data Preprocessing}\label{sec:prepro}

Neural data preprocessing is performed using the open-source plugin EEGLAB\cite{delorme2004eeglab} in the MATLAB software package. Initially, the raw EEG data is band-pass filtered using a zero-phase Hamming windowed sync FIR filter ranging between 0.1 and 40 Hz to eliminate artifacts in the low band (e.g., baseline drift and motion artifacts) and high band (e.g., electromyogram (EMG)). Subsequently, the filtered EEG data is re-referenced using common average referencing (CAR), followed by Independent Component Analysis (ICA) to eliminate artifacts related to electrooculography (EOG) and EMG. The number of removed independent components ranged from three to six across subjects, with the removed components primarily projecting over frontal cortical areas following the inverse ICA transform. Further, downsampling of denoised EEG data to 100 Hz sampling rate is performed for reduction of the computational cost. Eighteen EEG sensors around the motor-cortex region are selected for MKP.

Two key steps were involved in kinematic data preprocessing to eliminate measurement noise and signal scaling. First, a low-pass filter with a cutoff frequency of 2 Hz is applied to remove measurement noise from raw kinematics data. For scaling, min-max normalization is employed to scale the filtered data onto a range of [0, 1] as follows:

\begin{align}
K[t]=\frac{k[t]-{k}_{min}}{{k}_{max} - {k}_{min}}
\label{std1}
\end{align}
where,
$K[t]$ and $k[t]$ are the normalized and measured position coordinate values, respectively. ${k}_{max}$ and ${k}_{min}$ are maximum and minimum hand position coordinates, respectively. For scaling the kinematics data in the range of $[0, 1]$, min-max normalization is performed for the x, y, and z position coordinates. Further, the normalized kinematics data is downsampled from 500 Hz to 100 Hz sampling rate.


\subsection{EEG Source Imaging}\label{sec:esidata}
As the neural electrical signals transmit from the cortical surface to the scalp, they undergo distortion and dispersion due to volume-conduction effects. EEG source imaging (ESI) has the potential to mitigate cross-electrode correlation that results due to volume conduction effects. The fundamental principle behind ESI involves the estimation of primary cortical current activations based on the recorded scalp EEG signals. Two primary stages are involved in ESI: the solution of the forward problem and the inverse problem. The solution for the forward problem employs an anatomical head model to define the mapping between the voltages detected by EEG sensors on the scalp and the current activations originating from the cortical surface sources. Subsequently, the solution to the inverse problem aims to identify the cortical current distribution that aligns the scalp potential within a predefined set of constraints. It can be noted that scalp EEG data undergoes mapping onto a higher-dimensional cortical source grid with typical EEG sensors of around 30 and an estimated number of source dipoles of around 15,000. Due to the significant difference in the number of sensors and sources, ESI becomes a highly underdetermined problem. The two stages involved in the ESI are further detailed below:
\subsubsection{\textit{Forward Problem}} The computation of the lead field matrix is carried out in the forward modeling stage. The lead field matrix depicts the transmission of cortical source currents through various conductivity layers to the scalp electrodes. This procedure integrates the principles of EEG generation, incorporating considerations of Neumann and Dirichlet boundary conditions to establish the relationship between voltages recorded at the scalp and cortical current densities. The mathematical relationship can be expressed as:
\begin{equation*} \begin{bmatrix}\mathbf {E}\end{bmatrix}_{I\times N_{s}}=\begin{bmatrix}\mathbf {A}\end{bmatrix}_{I\times K}\,\begin{bmatrix}\mathbf {S}\end{bmatrix}_{K\times N_{s}} + \begin{bmatrix}\mathbf {\eta}\end{bmatrix}_{I\times N_{s}} \tag{1} \end{equation*}
where, $\mathbf{E}$ represents pre-processed scalp EEG data, $\mathbf{I}$ is the total number of EEG sensors (=32), $\mathbf{A}$ is the lead-field matrix, $\mathbf{S}$ denotes source activity, $\mathbf{K}$ is the total number of source dipoles, and $\mathbf{\eta}$ is the measured noise. 

The boundary element method (BEM) is utilized to compute the head model by employing the ICBM152 MRI template\cite{fonov2011unbiased} as the anatomy and subsequently co-registering with the positions of the EEG electrodes. The forward modeling is performed using the OpenMEEG \cite{gramfort2010openmeeg} with parameters $\sigma_{scalp} = 1$, $\sigma_{skull} = 0.0125$, and $\sigma_{brain} = 1$. In particular, the Brainstorm toolbox\cite{tadel2011brainstorm} is utilized for EEG source imaging (ESI).

\subsubsection{\textit{Inverse Problem}} The objective for solving the ESI inverse problem is to estimate the cortical source current signal $[S]$ using the lead field matrix $[A]$, derived from the head model computation. A noise perturbation matrix $[\eta]$ is introduced to account for errors resulting from solving the ill-posed and ill-conditioned inverse problem. In this study, the inverse problem is solved using the standard low-resolution electrical tomography (sLORETA) \cite{pascual2002standardized} method. sLORETA is utilized with normal to cortex dipole orientation and minimum norm imaging.

\begin{figure}[!h]
	\centering
	\includegraphics[width=0.30\textwidth]{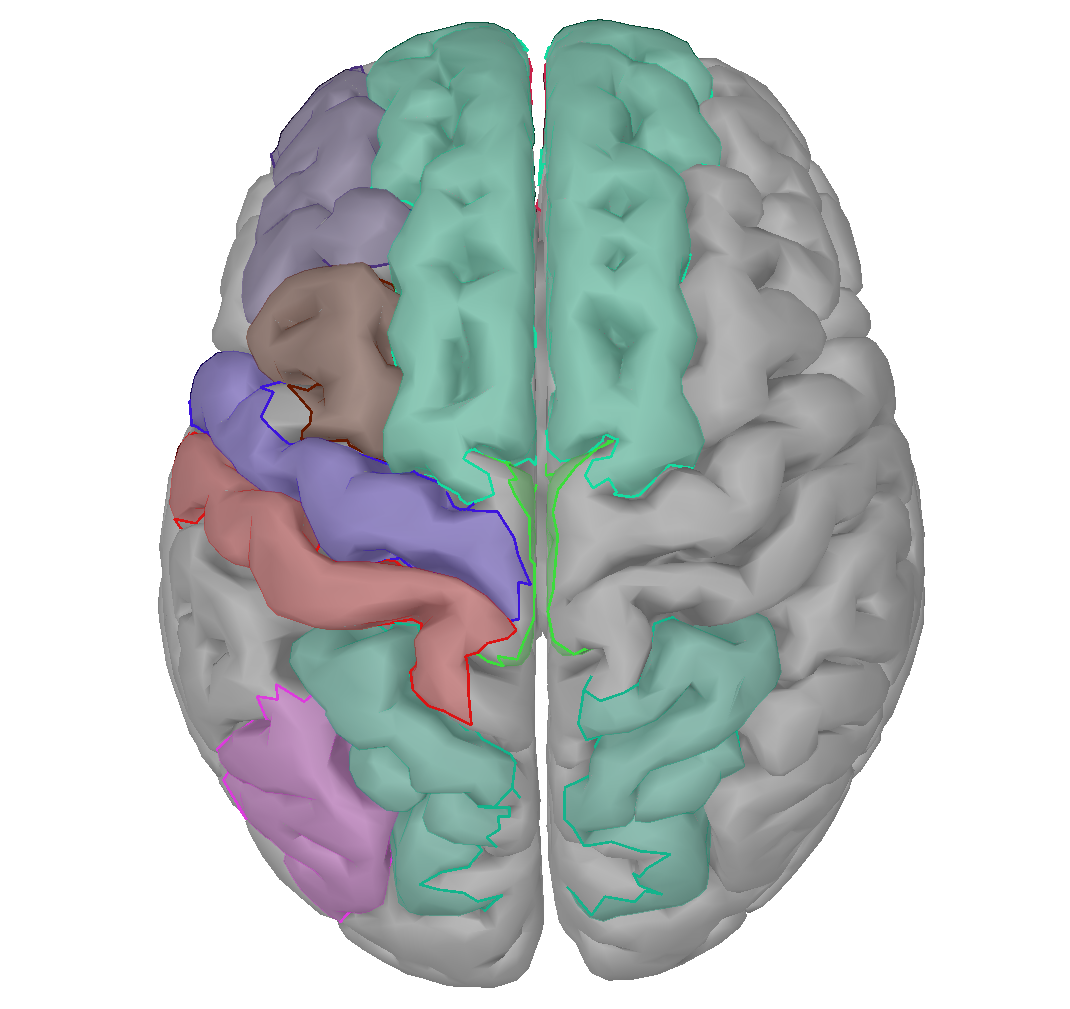}
        \centering
	\caption{Selected 15 Region of Interest (ROI) using Mindboggle Atlas}
	\label{04Mindboggle15ROI} 
\end{figure}

As a result of the transformation into source space, the number of signals increases to 15,000 source signals from 32 scalp EEG signals. The primary aim of regions of interest (ROI) scouting is to sub-group the number of signals within the source space based on different brain regions. The ROIs of the brain cortical area are defined based on the Mindboggle atlas\cite{klein2005mindboggle} for the ICBM152 MRI template. There are a total of 62 ROIs defined for the Mindboggle altas. However, even defined ROIs consist of signals in the order of hundreds. Hence, the mean of cortical source signals across each ROI for each timestamp is utilized. Further, ROIs are selected in the frontoparietal region of the brain. The selected ROIs are caudalmiddlefrontal (CMF), lateralorbitofrontal (LOF), medialorbitofrontal (MOFL), superiorfrontal (SF), paracentral (PCL), postcentral (PoCL), precentral (PreCL), superiorparietal (SP) and inferiorparietal (IP). In particular, a total of 15 ROIs were selected for the trajectory decoding, as shown in Fig. \ref{04Mindboggle15ROI}.
\subsection{Decoding Models}\label{sec:decoders}
Convolutional neural network (CNN) based Deep Learning architectures have achieved high accuracy for various EEG-based BCI applications. In this study, we have utilized modified versions of three popular architectures, EEGNet \cite{lawhern2018eegnet}, DeepConvNet \cite{schirrmeister2017deep}, and ShallowConvNet \cite{schirrmeister2017deep}, for continuous trajectory estimation. The modified versions of the models are depicted as rEEGNET, rDeepConvNet, and rShallowConvNet, where 'r' is the abbreviation for 'regression' here. The model architectures are summarized in tables \ref{tab:eegnet} to \ref{tab:scnet}; however, a comprehensive description is provided in the subsequent subsections:

\subsubsection{rEEGNet}\label{sec:eegnet} It is a CNN-based compact architecture employed across various BCI tasks. The compact nature of the decoder offers the advantage of efficient training with limited data. This study uses a modified version of EEGNet architecture for trajectory estimation. The model architecture can be segmented into two stages. In the first stage, a 2D convolutional (Conv) layer comprising 32 filters with kernel size (1, 32) is applied to the input signals. Afterward, a depthwise Conv layer is employed. The kernel size of the depthwise Conv layer is (18, 1) and (15, 1) for the sensor domain and source domain input features, respectively, due to the variation in input channels. Batch normalization is applied following both Conv layers. The linear activation function is utilized with the 2D CNN layer, while the Exponential Linear Unit (ELU) activation function is used for the depthwise CNN layer. The weights of the depthwise CNN layer are regularized with a maximum norm weight constraint of 1. A 2D average pooling layer with the size of (1, 2) is employed at the output stage along with a dropout layer with a rate of 0.5 to regularize the decoder. The second stage of the decoder included a separable Conv layer with 96 filters, each of size (1, 16), a batch normalization layer, and an activation function layer with ELU activation function. Subsequently, a 2D average pooling layer is applied with a size of (1, 4), and a feature vector is obtained by flattening the output of the average pooling layer. Further, the feature vector is fed to the output dense layer consisting of three neurons and the linear activation function. The incorporation of depthwise and separable Conv layers results in a significant reduction in the model parameter count, rendering a more compact model. Table \ref{tab:eegnet} shows the model architecture details for sensor domain input features.

\subsubsection{rDeepConvNet}\label{sec:dcnet} DeepConvNet model includes a total of four convolution blocks. The first block of the model consists of two 2D Conv layers followed by a batch normalization layer and an activation layer with the ELU activation function. The first Conv layer has the size of (1, 5); however, the kernel size of the second Conv layer is (18, 1) and (15, 1) for sensor domain and source domain input features, respectively. The remaining three blocks have identical configurations, with each block consisting of a 2D Conv layer of kernel size (1, 5), a batch normalization layer, and an activation layer with an ELU activation layer. The last block is followed by a 2D Max-pooling layer of kernel size and stride (1, 2). Each Conv layer consists of 25 filters in the first block; the number of filters in subsequent blocks is doubled, as shown in Table \ref{tab:dcnet}. The fourth block output is flattened and fed to the output dense layer with three neurons and linear activation.
\begin{table}[!t]
\centering
\caption{Model architecture of rEEGNet decoder for EEG sensor-domain features input.}
\scalebox{0.82}{
\centering
\begin{tabular}{cccc}
\hline
\multicolumn{1}{c}{\textbf{Layer}} & \multicolumn{1}{c}{\textbf{Kernel Size}} & \multicolumn{1}{c}{\textbf{\# of filters}} & \multicolumn{1}{c}{\textbf{Layer Parameters}}        \\ \hline \hline
Conv2D                               & (1, 32)                                   & 32                                          & Stride = (1, 1), Activation = Linear                  \\
BatchNorm                            & -                                         & -                                           & -                                                     \\
DepthwiseConv2D                      & (18, 1)                                   & -                                           & Depth multiplier = 3                                  \\
BatchNorm                            & -                                         & -                                           & -                                                     \\
Activation                           & -                                         & -                                           & Activation = ELU                                      \\
AveragePooling2D                     & (1, 2)                                    & -                                           & Stride = (1, 2)                                       \\
Dropout                              & -                                         & -                                           & Dropout Rate = 0.5                                    \\
SeperableConv2D                      & (1, 16)                                   & 96                                          & Zero Padding                                          \\
BatchNorm                            & -                                         & -                                           & -                                                     \\
Activation                           & -                                         & -                                           & Activation = ELU                                      \\
AveragePooling2D                     & (1, 4)                                    & -                                           & Stride = (1, 4)                                       \\
Dropout                              & -                                         & -                                           & Dropout Rate = 0.5                                    \\
Flatten                              & -                                         & -                                           & -                                                     \\ 
\multicolumn{1}{c}{Dense}          & \multicolumn{1}{c}{-}                    & \multicolumn{1}{c}{-}                      & \multicolumn{1}{c}{Neurons = 3, Activation = Linear} \\ \hline
\end{tabular}
}
\label{tab:eegnet}
\end{table}

\begin{table}[!t]
\centering
\caption{Model architecture of rDeepConvNet decoder for EEG sensor-domain features input.}
\scalebox{0.82}{
\centering
\begin{tabular}{cccc}
\hline
\multicolumn{1}{c}{\textbf{Layer}} & \multicolumn{1}{c}{\textbf{Kernel Size}} & \multicolumn{1}{c}{\textbf{\# of filters}} & \multicolumn{1}{c}{\textbf{Layer Parameters}}        \\ \hline \hline
Conv2D                               & (1, 5)                                    & 25                                          & Stride = (1, 1), Activation = Linear                  \\
Conv2D                               & (18, 1)                                   & 25                                          & Stride = (1, 1), Activation = Linear                  \\
BatchNorm                            & -                                         & -                                           & -                                                     \\
Activation                           & -                                         & -                                           & Activation = ELU                                      \\
Dropout                              & -                                         & -                                           & Dropout Rate = 0.5                                    \\
Conv2D                               & (1, 5)                                    & 50                                          & Stride = (1, 1), Activation = Linear                  \\
BatchNorm                            & -                                         & -                                           & -                                                     \\
Activation                           & -                                         & -                                           & Activation = ELU                                      \\
Dropout                              & -                                         & -                                           & Dropout Rate = 0.5                                    \\
Conv2D                               & (1, 5)                                    & 100                                         & Stride = (1, 1), Activation = Linear                  \\
BatchNorm                            & -                                         & -                                           & -                                                     \\
Activation                           & -                                         & -                                           & Activation = ELU                                      \\
Dropout                              & -                                         & -                                           & Dropout Rate = 0.5                                    \\
Conv2D                               & (1, 5)                                    & 200                                         & Stride = (1, 1), Activation = Linear                  \\
BatchNorm                            & -                                         & -                                           & -                                                     \\
Activation                           & -                                         & -                                           & Activation = ELU                                      \\
MaxPooling2D                         & (1, 2)                                    & -                                           & Stride = (1, 2)                                       \\
Dropout                              & -                                         & -                                           & Dropout Rate = 0.5                                    \\
Flatten                              & -                                         & -                                           & -                                                     \\ 
\multicolumn{1}{c}{Dense}          & \multicolumn{1}{c}{-}                    & \multicolumn{1}{c}{-}                      & \multicolumn{1}{c}{Neurons = 3, Activation = Linear} \\ \hline
\end{tabular}
}
\label{tab:dcnet}
\end{table}

\begin{table}[!t]
\centering
\caption{Model architecture of rShallowConvNet decoder for EEG sensor-domain features input.}
\scalebox{0.82}{
\centering
\begin{tabular}{cccc}
\hline
\multicolumn{1}{c}{\textbf{Layer}} & \multicolumn{1}{c}{\textbf{Kernel Size}} & \multicolumn{1}{c}{\textbf{\# of filters}} & \multicolumn{1}{c}{\textbf{Layer Parameters}}        \\ \hline \hline
Conv2D                               & (1, 13)                                   & 40                                          & Stride = (1, 1), Activation = Linear                  \\
Conv2D                               & (18, 1)                                   & 40                                          & Stride = (1, 1), Activation = Linear                  \\
BatchNorm                            & -                                         & -                                           & -                                                     \\
Activation                           & -                                         & -                                           & Activation = Square                                   \\
AveragePooling2D                     & (1, 5)                                    & -                                           & Stride = (1, 2)                                       \\
Activation                           & -                                         & -                                           & Activation = log                                      \\
Dropout                              & -                                         & -                                           & Dropout Rate = 0.5                                    \\
Flatten                              & -                                         & -                                           & -                                                     \\
\multicolumn{1}{c}{Dense}          & \multicolumn{1}{c}{-}                    & \multicolumn{1}{c}{-}                      & \multicolumn{1}{c}{Neurons = 3, Activation = Linear} \\ \hline
\end{tabular}
}
\label{tab:scnet}
\end{table}
\subsubsection{rShallowConvNet}\label{sec:scnet} The ShallowConvNet model consists of two 2D Conv layers with 40 filters and the linear activation function. The first 2D Conv layer has a kernel size of (1, 13), while the kernel size of the second Conv layer is (18, 1) and (15, 1) for sensor and source domain input features, respectively. The output of the Conv layers undergoes batch normalization, followed by the application of the square activation function. Subsequently, a 2D average-pooling layer of kernel size and stride (1, 5) and (1, 2), respectively, is utilized along with a logarithmic activation function. Further, the output of the average pooling layer is flattened to a vector and fed to the output-dense layer with three neurons with the linear activation function. A dropout layer with a drop rate of 0.5 is implemented to prevent overfitting.

\subsection{Experimental Details}\label{sec:expdetails}
The data normalization is performed on the EEG sensor and source time series using the z-score normalization technique, depicted as

\begin{align}
V^{n}[t]=\frac{v^{n}[t]-\Pi_{v_n}}{\rho_{v_n}}
\label{std2}
\end{align}
where, at time instant $t$, $v^{n}[t]$ and $V^{n}[t]$ are the $n^{th}$ channel denoised and normalized signal, respectively. $\Pi_{v_n}$ and $\rho_{v_n}$ denotes the mean and standard deviation of the $v^{n}$ signal, respectively.

Normalized sensor and source domain time series are utilized as input to neural decoders. For each trial, hand kinematics data is utilized from the movement initiation until the participant reinstated the hand to the initial resting position. Time-lagged EEG sensors and source time series segments that included the pre-motor neural information for hand trajectory decoding are utilized. This study considered various window sizes ($250$ $ms$ to $450$ $ms$) and lags ($50$ $ms$ to $200$ $ms$) with EEG data up to 650 ms prior to movement onset for the decoding analysis. The input to the neural decoders was a 2D matrix of dimension $C\times W$, where $C$ is the number of selected channels and $W$ is the window size. The number of selected channels ($C$) for the sensor and source domain time series are 18 and 15, respectively.

The dataset is divided into three data subsets for the performance evaluation of the neural decoders: training, validation, and test datasets. Model training is performed using the training dataset, while the validation dataset is utilized for tuning model hyper-parameters and avoiding over-fitting. The performance of the trained model was evaluated on the test dataset. Data corresponding to 294 trials is available for each participant in the WAY-EEG-GAL dataset. For intra-subject decoding analysis, the data for each participant is subdivided into training, validation, and test data corresponding to 234 trials, 30 trials, and 30 trials, respectively. For inter-subject decoding analysis, a 4-fold cross-validation strategy is adopted. Training and validation are carried out using data from nine subjects, and the trained model is evaluated using data corresponding to the remaining three subjects. The model training is performed using a mini-batch 
training process with a batch size of 64. The mean-squared error (mse) loss is minimized using the Adam optimization algorithm to define model parameters. The model undergoes training for a maximum of 400 epochs, and to prevent overfitting, early stopping is implemented based on the validation set mse loss, with a patience of 5 epochs.
\section{Results and Discussion}\label{sec:r&d}

\begin{table*}[t]
\centering
\caption{Mean PCC values for Intra-subject trajectory decoding in the x, y, and z directions using EEG sensor-domain and source-domain time series input. The effect on PCC values using different EEG lag and window sizes with different decoding models is also depicted.}
\scalebox{0.99}{
\centering
\begin{tabular}{|c|c|c|ccccc|ccccc|}
\hline
\multirow{3}{*}{\textbf{Direction}} & \multirow{3}{*}{\textbf{Decoders}} & \multirow{3}{*}{\textbf{EEG Lag}} & \multicolumn{5}{c|}{\textbf{Sensor Domain}}                                                                                                                    & \multicolumn{5}{c|}{\textbf{Source Domain}}                                                                                                                    \\ \cline{4-13} 
                                    &                                    &                                   & \multicolumn{5}{c|}{\textbf{EEG Window}}                                                                                                                       & \multicolumn{5}{c|}{\textbf{EEG Window}}                                                                                                                       \\ \cline{4-13} 
                                    &                                    &                                   & \multicolumn{1}{c|}{\textbf{250}} & \multicolumn{1}{c|}{\textbf{300}} & \multicolumn{1}{c|}{\textbf{350}} & \multicolumn{1}{c|}{\textbf{400}} & \textbf{450}   & \multicolumn{1}{c|}{\textbf{250}} & \multicolumn{1}{c|}{\textbf{300}} & \multicolumn{1}{c|}{\textbf{350}} & \multicolumn{1}{c|}{\textbf{400}} & \textbf{450}   \\ \hline \hline
\multirow{16}{*}{\textbf{x}}        & \multirow{4}{*}{\textbf{mLR}}      & \textbf{50}                       & \multicolumn{1}{c|}{0.340}        & \multicolumn{1}{c|}{0.351}        & \multicolumn{1}{c|}{0.369}        & \multicolumn{1}{c|}{0.374}        & 0.385          & \multicolumn{1}{c|}{0.410}        & \multicolumn{1}{c|}{0.420}        & \multicolumn{1}{c|}{0.429}        & \multicolumn{1}{c|}{0.429}        & 0.433          \\ \cline{3-13} 
                                    &                                    & \textbf{100}                      & \multicolumn{1}{c|}{0.333}        & \multicolumn{1}{c|}{0.353}        & \multicolumn{1}{c|}{0.361}        & \multicolumn{1}{c|}{0.374}        & 0.395          & \multicolumn{1}{c|}{0.402}        & \multicolumn{1}{c|}{0.414}        & \multicolumn{1}{c|}{0.417}        & \multicolumn{1}{c|}{0.423}        & 0.451          \\ \cline{3-13} 
                                    &                                    & \textbf{150}                      & \multicolumn{1}{c|}{0.335}        & \multicolumn{1}{c|}{0.346}        & \multicolumn{1}{c|}{0.361}        & \multicolumn{1}{c|}{0.383}        & 0.384          & \multicolumn{1}{c|}{0.396}        & \multicolumn{1}{c|}{0.402}        & \multicolumn{1}{c|}{0.411}        & \multicolumn{1}{c|}{0.441}        & 0.442          \\ \cline{3-13} 
                                    &                                    & \textbf{200}                      & \multicolumn{1}{c|}{0.329}        & \multicolumn{1}{c|}{0.346}        & \multicolumn{1}{c|}{0.369}        & \multicolumn{1}{c|}{0.372}        & 0.382          & \multicolumn{1}{c|}{0.384}        & \multicolumn{1}{c|}{0.396}        & \multicolumn{1}{c|}{0.429}        & \multicolumn{1}{c|}{0.431}        & 0.431          \\ \cline{2-13} 
                                    & \multirow{4}{*}{\textbf{rSCNet}}    & \textbf{50}                       & \multicolumn{1}{c|}{0.731}        & \multicolumn{1}{c|}{0.734}        & \multicolumn{1}{c|}{0.727}        & \multicolumn{1}{c|}{0.723}        & 0.714          & \multicolumn{1}{c|}{0.706}        & \multicolumn{1}{c|}{0.707}        & \multicolumn{1}{c|}{0.710}        & \multicolumn{1}{c|}{0.698}        & 0.692          \\ \cline{3-13} 
                                    &                                    & \textbf{100}                      & \multicolumn{1}{c|}{0.736}        & \multicolumn{1}{c|}{0.734}        & \multicolumn{1}{c|}{0.730}        & \multicolumn{1}{c|}{0.717}        & 0.723          & \multicolumn{1}{c|}{0.709}        & \multicolumn{1}{c|}{0.710}        & \multicolumn{1}{c|}{0.706}        & \multicolumn{1}{c|}{0.704}        & 0.700          \\ \cline{3-13} 
                                    &                                    & \textbf{150}                      & \multicolumn{1}{c|}{0.744}        & \multicolumn{1}{c|}{0.735}        & \multicolumn{1}{c|}{0.726}        & \multicolumn{1}{c|}{0.725}        & 0.717          & \multicolumn{1}{c|}{0.713}        & \multicolumn{1}{c|}{0.706}        & \multicolumn{1}{c|}{0.701}        & \multicolumn{1}{c|}{0.705}        & 0.699          \\ \cline{3-13} 
                                    &                                    & \textbf{200}                      & \multicolumn{1}{c|}{0.738}        & \multicolumn{1}{c|}{0.726}        & \multicolumn{1}{c|}{0.727}        & \multicolumn{1}{c|}{0.722}        & 0.715          & \multicolumn{1}{c|}{0.708}        & \multicolumn{1}{c|}{0.696}        & \multicolumn{1}{c|}{0.702}        & \multicolumn{1}{c|}{0.692}        & 0.683          \\ \cline{2-13} 
                                    & \multirow{4}{*}{\textbf{rDCNet}}    & \textbf{50}                       & \multicolumn{1}{c|}{0.741}        & \multicolumn{1}{c|}{0.752}        & \multicolumn{1}{c|}{0.745}        & \multicolumn{1}{c|}{0.715}        & 0.722          & \multicolumn{1}{c|}{0.717}        & \multicolumn{1}{c|}{0.732}        & \multicolumn{1}{c|}{0.731}        & \multicolumn{1}{c|}{0.718}        & 0.715          \\ \cline{3-13} 
                                    &                                    & \textbf{100}                      & \multicolumn{1}{c|}{0.746}        & \multicolumn{1}{c|}{0.752}        & \multicolumn{1}{c|}{0.741}        & \multicolumn{1}{c|}{0.730}        & 0.724          & \multicolumn{1}{c|}{0.722}        & \multicolumn{1}{c|}{0.723}        & \multicolumn{1}{c|}{0.722}        & \multicolumn{1}{c|}{0.707}        & 0.707          \\ \cline{3-13} 
                                    &                                    & \textbf{150}                      & \multicolumn{1}{c|}{0.752}        & \multicolumn{1}{c|}{0.741}        & \multicolumn{1}{c|}{0.744}        & \multicolumn{1}{c|}{0.739}        & 0.702          & \multicolumn{1}{c|}{0.722}        & \multicolumn{1}{c|}{0.730}        & \multicolumn{1}{c|}{0.717}        & \multicolumn{1}{c|}{0.715}        & 0.704          \\ \cline{3-13} 
                                    &                                    & \textbf{200}                      & \multicolumn{1}{c|}{0.734}        & \multicolumn{1}{c|}{0.743}        & \multicolumn{1}{c|}{0.743}        & \multicolumn{1}{c|}{0.733}        & 0.715          & \multicolumn{1}{c|}{0.716}        & \multicolumn{1}{c|}{0.718}        & \multicolumn{1}{c|}{0.730}        & \multicolumn{1}{c|}{0.708}        & 0.703          \\ \cline{2-13} 
                                    & \multirow{4}{*}{\textbf{rEEGNet}}   & \textbf{50}                       & \multicolumn{1}{c|}{0.781}        & \multicolumn{1}{c|}{0.783}        & \multicolumn{1}{c|}{0.787}        & \multicolumn{1}{c|}{0.785}        & 0.785          & \multicolumn{1}{c|}{0.753}        & \multicolumn{1}{c|}{0.766}        & \multicolumn{1}{c|}{0.769}        & \multicolumn{1}{c|}{0.767}        & 0.753          \\ \cline{3-13} 
                                    &                                    & \textbf{100}                      & \multicolumn{1}{c|}{0.785}        & \multicolumn{1}{c|}{0.781}        & \multicolumn{1}{c|}{0.785}        & \multicolumn{1}{c|}{0.781}        & \textbf{0.790} & \multicolumn{1}{c|}{0.761}        & \multicolumn{1}{c|}{0.768}        & \multicolumn{1}{c|}{0.758}        & \multicolumn{1}{c|}{0.759}        & \textbf{0.769} \\ \cline{3-13} 
                                    &                                    & \textbf{150}                      & \multicolumn{1}{c|}{0.781}        & \multicolumn{1}{c|}{0.781}        & \multicolumn{1}{c|}{0.780}        & \multicolumn{1}{c|}{0.788}        & 0.781          & \multicolumn{1}{c|}{0.758}        & \multicolumn{1}{c|}{0.758}        & \multicolumn{1}{c|}{0.755}        & \multicolumn{1}{c|}{0.767}        & 0.766          \\ \cline{3-13} 
                                    &                                    & \textbf{200}                      & \multicolumn{1}{c|}{0.783}        & \multicolumn{1}{c|}{0.777}        & \multicolumn{1}{c|}{0.787}        & \multicolumn{1}{c|}{0.779}        & 0.779          & \multicolumn{1}{c|}{0.760}        & \multicolumn{1}{c|}{0.751}        & \multicolumn{1}{c|}{0.765}        & \multicolumn{1}{c|}{0.759}        & 0.760          \\ \hline \hline 
\multirow{16}{*}{\textbf{y}}        & \multirow{4}{*}{\textbf{mLR}}      & \textbf{50}                       & \multicolumn{1}{c|}{0.349}        & \multicolumn{1}{c|}{0.358}        & \multicolumn{1}{c|}{0.377}        & \multicolumn{1}{c|}{0.379}        & 0.392          & \multicolumn{1}{c|}{0.416}        & \multicolumn{1}{c|}{0.427}        & \multicolumn{1}{c|}{0.436}        & \multicolumn{1}{c|}{0.434}        & 0.438          \\ \cline{3-13} 
                                    &                                    & \textbf{100}                      & \multicolumn{1}{c|}{0.340}        & \multicolumn{1}{c|}{0.361}        & \multicolumn{1}{c|}{0.366}        & \multicolumn{1}{c|}{0.382}        & 0.401          & \multicolumn{1}{c|}{0.407}        & \multicolumn{1}{c|}{0.420}        & \multicolumn{1}{c|}{0.422}        & \multicolumn{1}{c|}{0.429}        & 0.458          \\ \cline{3-13} 
                                    &                                    & \textbf{150}                      & \multicolumn{1}{c|}{0.342}        & \multicolumn{1}{c|}{0.351}        & \multicolumn{1}{c|}{0.369}        & \multicolumn{1}{c|}{0.390}        & 0.388          & \multicolumn{1}{c|}{0.401}        & \multicolumn{1}{c|}{0.407}        & \multicolumn{1}{c|}{0.417}        & \multicolumn{1}{c|}{0.448}        & 0.446          \\ \cline{3-13} 
                                    &                                    & \textbf{200}                      & \multicolumn{1}{c|}{0.333}        & \multicolumn{1}{c|}{0.354}        & \multicolumn{1}{c|}{0.376}        & \multicolumn{1}{c|}{0.377}        & 0.387          & \multicolumn{1}{c|}{0.388}        & \multicolumn{1}{c|}{0.403}        & \multicolumn{1}{c|}{0.436}        & \multicolumn{1}{c|}{0.436}        & 0.435          \\ \cline{2-13} 
                                    & \multirow{4}{*}{\textbf{rSCNet}}    & \textbf{50}                       & \multicolumn{1}{c|}{0.738}        & \multicolumn{1}{c|}{0.742}        & \multicolumn{1}{c|}{0.736}        & \multicolumn{1}{c|}{0.735}        & 0.727          & \multicolumn{1}{c|}{0.711}        & \multicolumn{1}{c|}{0.715}        & \multicolumn{1}{c|}{0.719}        & \multicolumn{1}{c|}{0.711}        & 0.708          \\ \cline{3-13} 
                                    &                                    & \textbf{100}                      & \multicolumn{1}{c|}{0.743}        & \multicolumn{1}{c|}{0.741}        & \multicolumn{1}{c|}{0.738}        & \multicolumn{1}{c|}{0.728}        & 0.733          & \multicolumn{1}{c|}{0.717}        & \multicolumn{1}{c|}{0.720}        & \multicolumn{1}{c|}{0.714}        & \multicolumn{1}{c|}{0.711}        & 0.709          \\ \cline{3-13} 
                                    &                                    & \textbf{150}                      & \multicolumn{1}{c|}{0.749}        & \multicolumn{1}{c|}{0.740}        & \multicolumn{1}{c|}{0.735}        & \multicolumn{1}{c|}{0.736}        & 0.729          & \multicolumn{1}{c|}{0.721}        & \multicolumn{1}{c|}{0.712}        & \multicolumn{1}{c|}{0.710}        & \multicolumn{1}{c|}{0.715}        & 0.706          \\ \cline{3-13} 
                                    &                                    & \textbf{200}                      & \multicolumn{1}{c|}{0.743}        & \multicolumn{1}{c|}{0.732}        & \multicolumn{1}{c|}{0.736}        & \multicolumn{1}{c|}{0.728}        & 0.725          & \multicolumn{1}{c|}{0.714}        & \multicolumn{1}{c|}{0.704}        & \multicolumn{1}{c|}{0.711}        & \multicolumn{1}{c|}{0.705}        & 0.697          \\ \cline{2-13} 
                                    & \multirow{4}{*}{\textbf{rDCNet}}    & \textbf{50}                       & \multicolumn{1}{c|}{0.753}        & \multicolumn{1}{c|}{0.763}        & \multicolumn{1}{c|}{0.762}        & \multicolumn{1}{c|}{0.751}        & 0.754          & \multicolumn{1}{c|}{0.727}        & \multicolumn{1}{c|}{0.741}        & \multicolumn{1}{c|}{0.742}        & \multicolumn{1}{c|}{0.734}        & 0.733          \\ \cline{3-13} 
                                    &                                    & \textbf{100}                      & \multicolumn{1}{c|}{0.754}        & \multicolumn{1}{c|}{0.761}        & \multicolumn{1}{c|}{0.760}        & \multicolumn{1}{c|}{0.754}        & 0.756          & \multicolumn{1}{c|}{0.733}        & \multicolumn{1}{c|}{0.735}        & \multicolumn{1}{c|}{0.731}        & \multicolumn{1}{c|}{0.730}        & 0.736          \\ \cline{3-13} 
                                    &                                    & \textbf{150}                      & \multicolumn{1}{c|}{0.760}        & \multicolumn{1}{c|}{0.751}        & \multicolumn{1}{c|}{0.757}        & \multicolumn{1}{c|}{0.755}        & 0.744          & \multicolumn{1}{c|}{0.731}        & \multicolumn{1}{c|}{0.738}        & \multicolumn{1}{c|}{0.729}        & \multicolumn{1}{c|}{0.736}        & 0.734          \\ \cline{3-13} 
                                    &                                    & \textbf{200}                      & \multicolumn{1}{c|}{0.743}        & \multicolumn{1}{c|}{0.752}        & \multicolumn{1}{c|}{0.755}        & \multicolumn{1}{c|}{0.753}        & 0.750          & \multicolumn{1}{c|}{0.725}        & \multicolumn{1}{c|}{0.725}        & \multicolumn{1}{c|}{0.739}        & \multicolumn{1}{c|}{0.730}        & 0.730          \\ \cline{2-13} 
                                    & \multirow{4}{*}{\textbf{rEEGNet}}   & \textbf{50}                       & \multicolumn{1}{c|}{0.787}        & \multicolumn{1}{c|}{0.788}        & \multicolumn{1}{c|}{0.792}        & \multicolumn{1}{c|}{0.787}        & 0.789          & \multicolumn{1}{c|}{0.759}        & \multicolumn{1}{c|}{0.774}        & \multicolumn{1}{c|}{0.775}        & \multicolumn{1}{c|}{0.773}        & 0.761          \\ \cline{3-13} 
                                    &                                    & \textbf{100}                      & \multicolumn{1}{c|}{0.789}        & \multicolumn{1}{c|}{0.788}        & \multicolumn{1}{c|}{0.788}        & \multicolumn{1}{c|}{0.786}        & \textbf{0.795} & \multicolumn{1}{c|}{0.768}        & \multicolumn{1}{c|}{0.774}        & \multicolumn{1}{c|}{0.763}        & \multicolumn{1}{c|}{0.768}        & \textbf{0.777} \\ \cline{3-13} 
                                    &                                    & \textbf{150}                      & \multicolumn{1}{c|}{0.786}        & \multicolumn{1}{c|}{0.783}        & \multicolumn{1}{c|}{0.782}        & \multicolumn{1}{c|}{0.793}        & 0.784          & \multicolumn{1}{c|}{0.764}        & \multicolumn{1}{c|}{0.764}        & \multicolumn{1}{c|}{0.762}        & \multicolumn{1}{c|}{0.776}        & 0.773          \\ \cline{3-13} 
                                    &                                    & \textbf{200}                      & \multicolumn{1}{c|}{0.787}        & \multicolumn{1}{c|}{0.780}        & \multicolumn{1}{c|}{0.791}        & \multicolumn{1}{c|}{0.785}        & 0.782          & \multicolumn{1}{c|}{0.766}        & \multicolumn{1}{c|}{0.756}        & \multicolumn{1}{c|}{0.771}        & \multicolumn{1}{c|}{0.767}        & 0.767          \\ \hline \hline
\multirow{16}{*}{\textbf{z}}        & \multirow{4}{*}{\textbf{mLR}}      & \textbf{50}                       & \multicolumn{1}{c|}{0.156}        & \multicolumn{1}{c|}{0.158}        & \multicolumn{1}{c|}{0.164}        & \multicolumn{1}{c|}{0.169}        & 0.173          & \multicolumn{1}{c|}{0.289}        & \multicolumn{1}{c|}{0.296}        & \multicolumn{1}{c|}{0.296}        & \multicolumn{1}{c|}{0.301}        & 0.304          \\ \cline{3-13} 
                                    &                                    & \textbf{100}                      & \multicolumn{1}{c|}{0.153}        & \multicolumn{1}{c|}{0.159}        & \multicolumn{1}{c|}{0.164}        & \multicolumn{1}{c|}{0.167}        & 0.173          & \multicolumn{1}{c|}{0.287}        & \multicolumn{1}{c|}{0.289}        & \multicolumn{1}{c|}{0.293}        & \multicolumn{1}{c|}{0.297}        & 0.298          \\ \cline{3-13} 
                                    &                                    & \textbf{150}                      & \multicolumn{1}{c|}{0.154}        & \multicolumn{1}{c|}{0.160}        & \multicolumn{1}{c|}{0.162}        & \multicolumn{1}{c|}{0.167}        & 0.173          & \multicolumn{1}{c|}{0.281}        & \multicolumn{1}{c|}{0.285}        & \multicolumn{1}{c|}{0.289}        & \multicolumn{1}{c|}{0.290}        & 0.290          \\ \cline{3-13} 
                                    &                                    & \textbf{200}                      & \multicolumn{1}{c|}{0.155}        & \multicolumn{1}{c|}{0.157}        & \multicolumn{1}{c|}{0.163}        & \multicolumn{1}{c|}{0.168}        & 0.170          & \multicolumn{1}{c|}{0.277}        & \multicolumn{1}{c|}{0.281}        & \multicolumn{1}{c|}{0.283}        & \multicolumn{1}{c|}{0.282}        & 0.287          \\ \cline{2-13} 
                                    & \multirow{4}{*}{\textbf{rSCNet}}    & \textbf{50}                       & \multicolumn{1}{c|}{0.562}        & \multicolumn{1}{c|}{0.565}        & \multicolumn{1}{c|}{0.564}        & \multicolumn{1}{c|}{0.560}        & 0.545          & \multicolumn{1}{c|}{0.568}        & \multicolumn{1}{c|}{0.578}        & \multicolumn{1}{c|}{0.580}        & \multicolumn{1}{c|}{0.582}        & 0.576          \\ \cline{3-13} 
                                    &                                    & \textbf{100}                      & \multicolumn{1}{c|}{0.559}        & \multicolumn{1}{c|}{0.562}        & \multicolumn{1}{c|}{0.555}        & \multicolumn{1}{c|}{0.555}        & 0.549          & \multicolumn{1}{c|}{0.576}        & \multicolumn{1}{c|}{0.579}        & \multicolumn{1}{c|}{0.578}        & \multicolumn{1}{c|}{0.574}        & 0.583          \\ \cline{3-13} 
                                    &                                    & \textbf{150}                      & \multicolumn{1}{c|}{0.565}        & \multicolumn{1}{c|}{0.558}        & \multicolumn{1}{c|}{0.550}        & \multicolumn{1}{c|}{0.559}        & 0.551          & \multicolumn{1}{c|}{0.584}        & \multicolumn{1}{c|}{0.576}        & \multicolumn{1}{c|}{0.569}        & \multicolumn{1}{c|}{0.581}        & 0.577          \\ \cline{3-13} 
                                    &                                    & \textbf{200}                      & \multicolumn{1}{c|}{0.553}        & \multicolumn{1}{c|}{0.549}        & \multicolumn{1}{c|}{0.553}        & \multicolumn{1}{c|}{0.543}        & 0.530          & \multicolumn{1}{c|}{0.575}        & \multicolumn{1}{c|}{0.571}        & \multicolumn{1}{c|}{0.581}        & \multicolumn{1}{c|}{0.572}        & 0.565          \\ \cline{2-13} 
                                    & \multirow{4}{*}{\textbf{rDCNet}}    & \textbf{50}                       & \multicolumn{1}{c|}{0.540}        & \multicolumn{1}{c|}{0.564}        & \multicolumn{1}{c|}{0.559}        & \multicolumn{1}{c|}{0.537}        & 0.549          & \multicolumn{1}{c|}{0.570}        & \multicolumn{1}{c|}{0.601}        & \multicolumn{1}{c|}{0.610}        & \multicolumn{1}{c|}{0.603}        & 0.596          \\ \cline{3-13} 
                                    &                                    & \textbf{100}                      & \multicolumn{1}{c|}{0.547}        & \multicolumn{1}{c|}{0.571}        & \multicolumn{1}{c|}{0.561}        & \multicolumn{1}{c|}{0.557}        & 0.542          & \multicolumn{1}{c|}{0.572}        & \multicolumn{1}{c|}{0.593}        & \multicolumn{1}{c|}{0.601}        & \multicolumn{1}{c|}{0.588}        & 0.597          \\ \cline{3-13} 
                                    &                                    & \textbf{150}                      & \multicolumn{1}{c|}{0.548}        & \multicolumn{1}{c|}{0.557}        & \multicolumn{1}{c|}{0.566}        & \multicolumn{1}{c|}{0.575}        & 0.520          & \multicolumn{1}{c|}{0.586}        & \multicolumn{1}{c|}{0.599}        & \multicolumn{1}{c|}{0.591}        & \multicolumn{1}{c|}{0.606}        & 0.587          \\ \cline{3-13} 
                                    &                                    & \textbf{200}                      & \multicolumn{1}{c|}{0.533}        & \multicolumn{1}{c|}{0.558}        & \multicolumn{1}{c|}{0.569}        & \multicolumn{1}{c|}{0.552}        & 0.529          & \multicolumn{1}{c|}{0.576}        & \multicolumn{1}{c|}{0.590}        & \multicolumn{1}{c|}{0.609}        & \multicolumn{1}{c|}{0.590}        & 0.592          \\ \cline{2-13} 
                                    & \multirow{4}{*}{\textbf{rEEGNet}}   & \textbf{50}                       & \multicolumn{1}{c|}{0.601}        & \multicolumn{1}{c|}{0.608}        & \multicolumn{1}{c|}{0.625}        & \multicolumn{1}{c|}{0.624}        & 0.624          & \multicolumn{1}{c|}{0.620}        & \multicolumn{1}{c|}{0.633}        & \multicolumn{1}{c|}{0.639}        & \multicolumn{1}{c|}{0.640}        & 0.635          \\ \cline{3-13} 
                                    &                                    & \textbf{100}                      & \multicolumn{1}{c|}{0.597}        & \multicolumn{1}{c|}{0.607}        & \multicolumn{1}{c|}{0.617}        & \multicolumn{1}{c|}{0.616}        & \textbf{0.637} & \multicolumn{1}{c|}{0.627}        & \multicolumn{1}{c|}{0.632}        & \multicolumn{1}{c|}{0.633}        & \multicolumn{1}{c|}{0.634}        & \textbf{0.647} \\ \cline{3-13} 
                                    &                                    & \textbf{150}                      & \multicolumn{1}{c|}{0.604}        & \multicolumn{1}{c|}{0.598}        & \multicolumn{1}{c|}{0.609}        & \multicolumn{1}{c|}{0.626}        & 0.624          & \multicolumn{1}{c|}{0.616}        & \multicolumn{1}{c|}{0.626}        & \multicolumn{1}{c|}{0.621}        & \multicolumn{1}{c|}{0.645}        & 0.640          \\ \cline{3-13} 
                                    &                                    & \textbf{200}                      & \multicolumn{1}{c|}{0.599}        & \multicolumn{1}{c|}{0.601}        & \multicolumn{1}{c|}{0.617}        & \multicolumn{1}{c|}{0.611}        & 0.616          & \multicolumn{1}{c|}{0.620}        & \multicolumn{1}{c|}{0.613}        & \multicolumn{1}{c|}{0.639}        & \multicolumn{1}{c|}{0.637}        & 0.639          \\ \hline
\end{tabular}
}
\label{tab:intrasub_pccresults}
\vspace{0.15cm}
\scriptsize{\\
Note: the bold entries represent the highest PCC value obtained in the x, y, and z directions for the EEG sensor-domain and source-domain input features.\\
rDCNet - rDeepConvNet, rSCNet - rShallowConvNet}
\end{table*}

\subsection{Performance Metric}\label{sec:metric}
The Pearson correlation coefficient (PCC) is employed as a performance metric for evaluating the MKP efficacy of the decoding models. This coefficient is calculated across measured and predicted kinematics trajectory, representing linear correlation with output values in the range $[-1,1]$. A PCC value of -1 and +1 indicates a strong negative and positive correlation, respectively, while a PCC value of 0 represents no correlation. The mathematical expression that represents the Pearson correlation coefficient between the measured ($C_x$) and estimated ($C_y$) is depicted as follows:

\begin{align}
	\Pi (C_x,C_y)=\frac{1}{T-1}\sum_{i=1}^{T}\left ( \frac{C_x^i-\alpha^{C_x}}{\beta^{C_x}} \right )\left( \frac{C_y^i-\alpha^{C_y}}{\beta^{C_y}} \right )
	\label{corr}
\end{align}
where, $\alpha_m$ and $\beta_m$ are the mean and standard deviation of $m$, respectively, with $m\in \{C_x,C_y\}$.

 
\subsection{Results}\label{sec:results}
In the first set of experiments, the performance of motor kinematics prediction is explored for intra-subject settings using sensor-domain and source-domain EEG features. Table \ref{tab:intrasub_pccresults} presents the mean correlation values for twelve subjects using different decoding models. The decoding analysis is performed for hand kinematics trajectory in the x, y, and z-directions. In addition, the effect of EEG lag and window size is also presented in Table \ref{tab:intrasub_pccresults}. EEG lag up to $200$ $ms$ is analyzed with various EEG window sizes in the range of $250-450$ $ms$.

Furthermore, the performance analysis for inter-subject decoding is also explored. The mean correlation values across twelve subjects using the mLR model and three deep learning models are presented in Table \ref{tab:intersub_pccresults}. The decoding analysis is performed for $50$ $ms$ time lag and window size up to $450$ $ms$ for the MKP of hand position in x, y, and z-directions. In particular, sensor-domain and source-domain EEG features are utilized for comparing MKP performance.
\begin{table*}[t]
\centering
\caption{Mean PCC values for Inter-subject trajectory decoding in the x, y, and z directions using EEG sensor-domain and source-domain time series input. The effect on PCC values using 50 ms lag and different window sizes with different decoding models is also depicted.}
\scalebox{0.99}{
\centering
\begin{tabular}{|c|c|lllll|lllll|}
\hline
\multirow{3}{*}{\textbf{Direction}} & \multirow{3}{*}{\textbf{Decoders}} & \multicolumn{5}{c|}{\textbf{Sensor Domain}}                                                                                                                                         & \multicolumn{5}{c|}{\textbf{Source Domain}}                                                                                                                                           \\ \cline{3-12} 
                                    &                                    & \multicolumn{5}{c|}{\textbf{Window size (ms)}}                                                                                                                                      & \multicolumn{5}{c|}{\textbf{Window size (ms)}}                                                                                                                                        \\ \cline{3-12} 
                                    &                                    & \multicolumn{1}{c|}{\textbf{250}} & \multicolumn{1}{c|}{\textbf{300}} & \multicolumn{1}{c|}{\textbf{350}} & \multicolumn{1}{c|}{\textbf{400}}   & \multicolumn{1}{c|}{\textbf{450}} & \multicolumn{1}{c|}{\textbf{250}} & \multicolumn{1}{c|}{\textbf{300}} & \multicolumn{1}{c|}{\textbf{350}}   & \multicolumn{1}{c|}{\textbf{400}}   & \multicolumn{1}{c|}{\textbf{450}} \\ \hline
\multirow{4}{*}{\textbf{x}}         & \textbf{mLR}                       & \multicolumn{1}{l|}{0.143}        & \multicolumn{1}{l|}{0.151}        & \multicolumn{1}{l|}{0.168}        & \multicolumn{1}{l|}{0.186}          & 0.203                             & \multicolumn{1}{l|}{0.150}        & \multicolumn{1}{l|}{0.150}        & \multicolumn{1}{l|}{0.167}          & \multicolumn{1}{l|}{0.182}          & 0.201                             \\ \cline{2-12} 
                                    & \textbf{rSCNet}                     & \multicolumn{1}{l|}{0.685}        & \multicolumn{1}{l|}{0.692}        & \multicolumn{1}{l|}{0.691}        & \multicolumn{1}{l|}{0.694}          & 0.692                             & \multicolumn{1}{l|}{0.625}        & \multicolumn{1}{l|}{0.625}        & \multicolumn{1}{l|}{0.627}          & \multicolumn{1}{l|}{0.621}          & 0.633                             \\ \cline{2-12} 
                                    & \textbf{rDCNet}                     & \multicolumn{1}{l|}{0.730}        & \multicolumn{1}{l|}{0.731}        & \multicolumn{1}{l|}{0.731}        & \multicolumn{1}{l|}{0.731}          & 0.715                             & \multicolumn{1}{l|}{0.664}        & \multicolumn{1}{l|}{0.658}        & \multicolumn{1}{l|}{0.658}          & \multicolumn{1}{l|}{0.652}          & 0.643                             \\ \cline{2-12} 
                                    & \textbf{rEEGNet}                    & \multicolumn{1}{l|}{0.745}        & \multicolumn{1}{l|}{0.744}        & \multicolumn{1}{l|}{0.749}        & \multicolumn{1}{l|}{\textbf{0.753}} & 0.748                             & \multicolumn{1}{l|}{0.683}        & \multicolumn{1}{l|}{0.682}        & \multicolumn{1}{l|}{\textbf{0.684}} & \multicolumn{1}{l|}{0.683}          & 0.683                             \\ \hline \hline
\multirow{4}{*}{\textbf{y}}         & \textbf{mLR}                       & \multicolumn{1}{l|}{0.145}        & \multicolumn{1}{l|}{0.156}        & \multicolumn{1}{l|}{0.172}        & \multicolumn{1}{l|}{0.188}          & 0.203                             & \multicolumn{1}{l|}{0.158}        & \multicolumn{1}{l|}{0.160}        & \multicolumn{1}{l|}{0.176}          & \multicolumn{1}{l|}{0.188}          & 0.204                             \\ \cline{2-12} 
                                    & \textbf{rSCNet}                     & \multicolumn{1}{l|}{0.688}        & \multicolumn{1}{l|}{0.694}        & \multicolumn{1}{l|}{0.697}        & \multicolumn{1}{l|}{0.703}          & 0.695                             & \multicolumn{1}{l|}{0.627}        & \multicolumn{1}{l|}{0.629}        & \multicolumn{1}{l|}{0.628}          & \multicolumn{1}{l|}{0.621}          & 0.636                             \\ \cline{2-12} 
                                    & \textbf{rDCNet}                     & \multicolumn{1}{l|}{0.729}        & \multicolumn{1}{l|}{0.731}        & \multicolumn{1}{l|}{0.731}        & \multicolumn{1}{l|}{0.733}          & 0.714                             & \multicolumn{1}{l|}{0.668}        & \multicolumn{1}{l|}{0.664}        & \multicolumn{1}{l|}{0.660}          & \multicolumn{1}{l|}{0.656}          & 0.643                             \\ \cline{2-12} 
                                    & \textbf{rEEGNet}                    & \multicolumn{1}{l|}{0.745}        & \multicolumn{1}{l|}{0.743}        & \multicolumn{1}{l|}{0.750}        & \multicolumn{1}{l|}{\textbf{0.756}} & 0.744                             & \multicolumn{1}{l|}{0.685}        & \multicolumn{1}{l|}{0.686}        & \multicolumn{1}{l|}{\textbf{0.687}} & \multicolumn{1}{l|}{\textbf{0.687}} & 0.680                             \\ \hline \hline
\multirow{4}{*}{\textbf{z}}         & \textbf{mLR}                       & \multicolumn{1}{l|}{0.051}        & \multicolumn{1}{l|}{0.051}        & \multicolumn{1}{l|}{0.048}        & \multicolumn{1}{l|}{0.047}          & 0.047                             & \multicolumn{1}{l|}{0.146}        & \multicolumn{1}{l|}{0.144}        & \multicolumn{1}{l|}{0.142}          & \multicolumn{1}{l|}{0.147}          & 0.149                             \\ \cline{2-12} 
                                    & \textbf{rSCNet}                     & \multicolumn{1}{l|}{0.412}        & \multicolumn{1}{l|}{0.407}        & \multicolumn{1}{l|}{0.391}        & \multicolumn{1}{l|}{0.402}          & 0.419                             & \multicolumn{1}{l|}{0.431}        & \multicolumn{1}{l|}{0.442}        & \multicolumn{1}{l|}{0.448}          & \multicolumn{1}{l|}{0.445}          & 0.453                             \\ \cline{2-12} 
                                    & \textbf{rDCNet}                     & \multicolumn{1}{l|}{0.442}        & \multicolumn{1}{l|}{0.445}        & \multicolumn{1}{l|}{0.455}        & \multicolumn{1}{l|}{0.468}          & 0.465                             & \multicolumn{1}{l|}{0.468}        & \multicolumn{1}{l|}{0.473}        & \multicolumn{1}{l|}{0.491}          & \multicolumn{1}{l|}{0.497}          & 0.493                             \\ \cline{2-12} 
                                    & \textbf{rEEGNet}                    & \multicolumn{1}{l|}{0.465}        & \multicolumn{1}{l|}{0.454}        & \multicolumn{1}{l|}{0.461}        & \multicolumn{1}{l|}{0.478}          & \textbf{0.490}                    & \multicolumn{1}{l|}{0.484}        & \multicolumn{1}{l|}{0.495}        & \multicolumn{1}{l|}{\textbf{0.515}} & \multicolumn{1}{l|}{0.506}          & 0.509                             \\ \hline
\end{tabular}
}
\label{tab:intersub_pccresults}
\vspace{0.15cm}
\scriptsize{\\
Note: the bold entries represent the highest PCC value obtained in the x, y, and z directions for the EEG sensor-domain and source-domain input features.\\
rDCNet - rDeepConvNet, rSCNet - rShallowConvNet}
\end{table*}
\subsection{Discussion}\label{sec:discussion}

\subsubsection{mLR Performance analysis}
In this work, the mLR model is utilized for MKP using EEG sensor-domain and source-domain features for hand trajectory in x, y, and z-directions. The decoding performance is evaluated for intra-subject and inter-subject decoding settings as shown in Table \ref{tab:intrasub_pccresults} and \ref{tab:intersub_pccresults}, respectively. In intra-subject decoding analysis, the mLR with source-domain EEG features has better decoding performance in x, y, and z-directions in comparison with sensor-domain-based MKP. The same can be depicted from $p$-values shown in Table \ref{tab:sensor_source_p_values}. However, the decoding performance degrades in the case of inter-subject decoding. Also, the decoding performance is statistically similar using the EEG sensor-domain and source-domain features for MKP, as shown in Table \ref{tab:sensor_source_p_values}. Furthermore, the deep learning-based decoders outperformed the mLR model with sensor-domain and source-domain features in intra-subject and inter-subject settings, as depicted from the $p$-values shown in Table \ref{tab:intrasub_p_values}-\ref{tab:sensor_source_p_values}.

\subsubsection{Deep Learning Model Analysis}
Three deep learning models, namely, rShallowConvNet, rDeepConvNet, and rEEGNet, are utilized to access the MKP performance during the grasp-and-lift task. As evident from the results in Tables \ref{tab:intrasub_pccresults} - \ref{tab:intersub_pccresults}, the rEEGNet decoding model outperforms the models for MKP in the x, y, and z-directions. The model architecture consists of depthwise and separable convolution layers that are designed for feature extraction from EEG data within end-to-end model training. Further, rEEGNet has the minimum parameters to train among the deep learning models utilized for MKP analysis. Subsequently, a detailed $t$-test is performed to compare the trajectory decoding performance of the various decoding models. The statistical analysis is performed to compare the decoding performance of models with sensor-domain features and source-domain input features. The results of the $t$-test are shown in Table \ref{tab:intrasub_p_values} for the intra-subject decoding case. It can be noted that the deep learning-based decoders performed significantly better than the traditionally used mLR model. In particular, the rEEGNet model decoding performance is statistically better in comparison to other decoding models. For the inter-subject case, all p-values are $<0.05$ in the $t$-test, which signifies the superior performance of the rEEGNet model over the other decoding models.

\subsubsection{EEG Lag Analysis}
In this analysis, hand trajectory is decoded using pre-movement EEG data with distinct window sizes and lags. For intra-subject decoding analysis, EEG segment up to $200$ $ms$ prior to movement-onset and size in the $250 - 450$ $ms$ range is utilized as shown in Table \ref{tab:intrasub_pccresults}. The best correlation values are obtained for an EEG lag of $100$ $ms$ and window size of $450$ $ms$ while using the rEEGNet decoding model. For sensor-domain EEG features, the mean correlation values obtained are $0.790,$ $0.795,$ and $0.637$ for x, y, and z-directions, respectively, while the mean correlation values of $0.769,$ $0.777,$ and $0.647$ are observed with EEG source-domain features. In the case of inter-subject decoding analysis, EEG segments with varying sizes and $50$ $ms$ pre-movement EEG data are taken as depicted in Table \ref{tab:intersub_pccresults}. The rEEGNet model has the best decoding performance with sensor-domain as well as source-domain input features. With sensor-domain input features, the best mean correlation values of $0.753$ and $0.756$ are observed with $400$ $ms$ window size in the x and y directions, respectively, while $0.490$ mean correlation value is obtained with $450$ $ms$ window size in the z-direction. The decoding performance with source-domain features is observed with $350$ $ms$ window size and mean correlation values of $0.684,$ $0.687,$ and $0.515$ in the x, y, and z-directions, respectively.
\begin{table*}
\centering
\caption{$p$ values of one-tailed $t$-test between sensor-domain and source-domain features-based trajectory decoders for Intra-subject and Inter-subject settings in the x, y, and z directions.}
\scalebox{0.93}{
\begin{tabular}{ll}
\begin{tabular}{cccccc|}
\hline \hline
\multicolumn{6}{c}{Intra-Subject}                                                                                                                                                              \\ \hline \hline
\multicolumn{2}{|c|}{\multirow{2}{*}{x-direction}}                                    & \multicolumn{4}{c|}{Source Domain}                                                                       \\ \cline{3-6} 
\multicolumn{2}{|c|}{}                                                                & \multicolumn{1}{c|}{mLR}      & \multicolumn{1}{c|}{rSCNet}   & \multicolumn{1}{c|}{rDCNet}   & rEEGNet  \\ \hline
\multicolumn{1}{|c|}{\multirow{4}{*}{\rotatebox[origin=c]{90}{Sensor}\vspace{0.1cm} \rotatebox[origin=c]{90}{Domain}}} & \multicolumn{1}{c|}{mLR}     & \multicolumn{1}{c|}{7.40×10$^{-5}$} & \multicolumn{1}{c|}{2.93×10$^{-6}$} & \multicolumn{1}{c|}{1.58×10$^{-6}$} & 6.45×10$^{-7}$ \\ \cline{2-6} 
\multicolumn{1}{|c|}{}                                 & \multicolumn{1}{c|}{rSCNet}  & \multicolumn{1}{c|}{9.35×10$^{-7}$} & \multicolumn{1}{c|}{1.14×10$^{-4}$} & \multicolumn{1}{c|}{2.09×10$^{-1}$} & 4.51×10$^{-4}$ \\ \cline{2-6} 
\multicolumn{1}{|c|}{}                                 & \multicolumn{1}{c|}{rDCNet}  & \multicolumn{1}{c|}{3.03×10$^{-6}$} & \multicolumn{1}{c|}{9.98×10$^{-4}$} & \multicolumn{1}{c|}{3.09×10$^{-2}$} & 1.08×10$^{-2}$ \\ \cline{2-6} 
\multicolumn{1}{|c|}{}                                 & \multicolumn{1}{c|}{rEEGNet} & \multicolumn{1}{c|}{1.76×10$^{-8}$} & \multicolumn{1}{c|}{1.18×10$^{-5}$} & \multicolumn{1}{c|}{3.37×10$^{-5}$} & 1.08×10$^{-3}$ \\ \hline \hline
\multicolumn{2}{|c|}{\multirow{2}{*}{y-direction}}                                    & \multicolumn{4}{c|}{Source Domain}                                                                       \\ \cline{3-6} 
\multicolumn{2}{|c|}{}                                                                & \multicolumn{1}{c|}{mLR}      & \multicolumn{1}{c|}{rSCNet}   & \multicolumn{1}{c|}{rDCNet}   & rEEGNet  \\ \hline
\multicolumn{1}{|c|}{\multirow{4}{*}{\rotatebox[origin=c]{90}{Sensor}\vspace{0.1cm} \rotatebox[origin=c]{90}{Domain}}} & \multicolumn{1}{c|}{mLR}     & \multicolumn{1}{c|}{6.87×10$^{-5}$} & \multicolumn{1}{c|}{1.11×10$^{-6}$} & \multicolumn{1}{c|}{6.08×10$^{-7}$} & 5.56×10$^{-7}$ \\ \cline{2-6} 
\multicolumn{1}{|c|}{}                                 & \multicolumn{1}{c|}{rSCNet}  & \multicolumn{1}{c|}{3.81×10$^{-7}$} & \multicolumn{1}{c|}{1.47×10$^{-4}$} & \multicolumn{1}{c|}{4.94×10$^{-1}$} & 2.19×10$^{-4}$ \\ \cline{2-6} 
\multicolumn{1}{|c|}{}                                 & \multicolumn{1}{c|}{rDCNet}  & \multicolumn{1}{c|}{1.25×10$^{-7}$} & \multicolumn{1}{c|}{4.71×10$^{-6}$} & \multicolumn{1}{c|}{7.12×10$^{-5}$} & 7.36×10$^{-3}$ \\ \cline{2-6} 
\multicolumn{1}{|c|}{}                                 & \multicolumn{1}{c|}{rEEGNet} & \multicolumn{1}{c|}{3.04×10$^{-8}$} & \multicolumn{1}{c|}{5.20×10$^{-7}$} & \multicolumn{1}{c|}{1.29×10$^{-5}$} & 1.69×10$^{-3}$ \\ \hline \hline
\multicolumn{2}{|c|}{\multirow{2}{*}{z-direction}}                                    & \multicolumn{4}{c|}{Source Domain}                                                                       \\ \cline{3-6} 
\multicolumn{2}{|c|}{}                                                                & \multicolumn{1}{c|}{mLR}      & \multicolumn{1}{c|}{rSCNet}   & \multicolumn{1}{c|}{rDCNet}   & rEEGNet  \\ \hline
\multicolumn{1}{|c|}{\multirow{4}{*}{\rotatebox[origin=c]{90}{Sensor}\vspace{0.1cm} \rotatebox[origin=c]{90}{Domain}}} & \multicolumn{1}{c|}{mLR}     & \multicolumn{1}{c|}{1.36×10$^{-8}$} & \multicolumn{1}{c|}{5.27×10$^{-9}$} & \multicolumn{1}{c|}{1.07×10$^{-7}$} & 3.12×10$^{-9}$ \\ \cline{2-6} 
\multicolumn{1}{|c|}{}                                 & \multicolumn{1}{c|}{rSCNet}  & \multicolumn{1}{c|}{6.54×10$^{-7}$} & \multicolumn{1}{c|}{6.59×10$^{-3}$} & \multicolumn{1}{c|}{4.19×10$^{-3}$} & 7.17×10$^{-5}$ \\ \cline{2-6} 
\multicolumn{1}{|c|}{}                                 & \multicolumn{1}{c|}{rDCNet}  & \multicolumn{1}{c|}{8.18×10$^{-7}$} & \multicolumn{1}{c|}{3.18×10$^{-3}$} & \multicolumn{1}{c|}{8.72×10$^{-4}$} & 5.79×10$^{-5}$ \\ \cline{2-6} 
\multicolumn{1}{|c|}{}                                 & \multicolumn{1}{c|}{rEEGNet} & \multicolumn{1}{c|}{2.93×10$^{-8}$} & \multicolumn{1}{c|}{1.39×10$^{-4}$} & \multicolumn{1}{c|}{4.68×10$^{-3}$} & 8.93×10$^{-4}$ \\ \hline
\end{tabular}
&
\begin{tabular}{cccccc|}
\hline \hline
\multicolumn{6}{c}{Inter-Subject}                                                                                                                                                              \\ \hline \hline
\multicolumn{2}{|c|}{\multirow{2}{*}{x-direction}}                                    & \multicolumn{4}{c|}{Source Domain}                                                                       \\ \cline{3-6} 
\multicolumn{2}{|c|}{}                                                                & \multicolumn{1}{c|}{mLR}      & \multicolumn{1}{c|}{rSCNet}   & \multicolumn{1}{c|}{rDCNet}   & rEEGNet  \\ \hline
\multicolumn{1}{|c|}{\multirow{4}{*}{\rotatebox[origin=c]{90}{Sensor}\vspace{0.1cm} \rotatebox[origin=c]{90}{Domain}}} & \multicolumn{1}{c|}{mLR}     & \multicolumn{1}{c|}{4.32×10$^{-1}$} & \multicolumn{1}{c|}{7.54×10$^{-7}$} & \multicolumn{1}{c|}{2.32×10$^{-6}$} & 5.82×10$^{-7}$ \\ \cline{2-6} 
\multicolumn{1}{|c|}{}                                 & \multicolumn{1}{c|}{rSCNet}  & \multicolumn{1}{c|}{2.64×10$^{-7}$} & \multicolumn{1}{c|}{5.80×10$^{-6}$} & \multicolumn{1}{c|}{7.59×10$^{-4}$} & 2.48×10$^{-3}$ \\ \cline{2-6} 
\multicolumn{1}{|c|}{}                                 & \multicolumn{1}{c|}{rDCNet}  & \multicolumn{1}{c|}{6.96×10$^{-7}$} & \multicolumn{1}{c|}{1.55×10$^{-5}$} & \multicolumn{1}{c|}{1.75×10$^{-6}$} & 6.57×10$^{-5}$ \\ \cline{2-6} 
\multicolumn{1}{|c|}{}                                 & \multicolumn{1}{c|}{rEEGNet} & \multicolumn{1}{c|}{1.57×10$^{-7}$} & \multicolumn{1}{c|}{1.01×10$^{-6}$} & \multicolumn{1}{c|}{1.75×10$^{-5}$} & 6.85×10$^{-7}$ \\ \hline \hline
\multicolumn{2}{|c|}{\multirow{2}{*}{y-direction}}                                    & \multicolumn{4}{c|}{Source Domain}                                                                       \\ \cline{3-6} 
\multicolumn{2}{|c|}{}                                                                & \multicolumn{1}{c|}{mLR}      & \multicolumn{1}{c|}{rSCNet}   & \multicolumn{1}{c|}{rDCNet}   & rEEGNet  \\ \hline
\multicolumn{1}{|c|}{\multirow{4}{*}{\rotatebox[origin=c]{90}{Sensor}\vspace{0.1cm} \rotatebox[origin=c]{90}{Domain}}} & \multicolumn{1}{c|}{mLR}     & \multicolumn{1}{c|}{6.22×10$^{-2}$} & \multicolumn{1}{c|}{7.11×10$^{-7}$} & \multicolumn{1}{c|}{2.35×10$^{-6}$} & 6.30×10$^{-7}$ \\ \cline{2-6} 
\multicolumn{1}{|c|}{}                                 & \multicolumn{1}{c|}{rSCNet}  & \multicolumn{1}{c|}{1.32×10$^{-7}$} & \multicolumn{1}{c|}{3.88×10$^{-5}$} & \multicolumn{1}{c|}{1.48×10$^{-3}$} & 5.66×10$^{-3}$ \\ \cline{2-6} 
\multicolumn{1}{|c|}{}                                 & \multicolumn{1}{c|}{rDCNet}  & \multicolumn{1}{c|}{5.22×10$^{-7}$} & \multicolumn{1}{c|}{2.82×10$^{-5}$} & \multicolumn{1}{c|}{5.41×10$^{-6}$} & 1.97×10$^{-5}$ \\ \cline{2-6} 
\multicolumn{1}{|c|}{}                                 & \multicolumn{1}{c|}{rEEGNet} & \multicolumn{1}{c|}{1.33×10$^{-7}$} & \multicolumn{1}{c|}{6.51×10$^{-6}$} & \multicolumn{1}{c|}{2.94×10$^{-5}$} & 2.92×10$^{-6}$ \\ \hline \hline
\multicolumn{2}{|c|}{\multirow{2}{*}{z-direction}}                                    & \multicolumn{4}{c|}{Source Domain}                                                                       \\ \cline{3-6} 
\multicolumn{2}{|c|}{}                                                                & \multicolumn{1}{c|}{mLR}      & \multicolumn{1}{c|}{rSCNet}   & \multicolumn{1}{c|}{rDCNet}   & rEEGNet  \\ \hline
\multicolumn{1}{|c|}{\multirow{4}{*}{\rotatebox[origin=c]{90}{Sensor}\vspace{0.1cm} \rotatebox[origin=c]{90}{Domain}}} & \multicolumn{1}{c|}{mLR}     & \multicolumn{1}{c|}{3.64×10$^{-7}$} & \multicolumn{1}{c|}{4.96×10$^{-8}$} & \multicolumn{1}{c|}{1.79×10$^{-7}$} & 1.22×10$^{-7}$ \\ \cline{2-6} 
\multicolumn{1}{|c|}{}                                 & \multicolumn{1}{c|}{rSCNet}  & \multicolumn{1}{c|}{1.40×10$^{-7}$} & \multicolumn{1}{c|}{1.96×10$^{-3}$} & \multicolumn{1}{c|}{3.88×10$^{-4}$} & 2.14×10$^{-4}$ \\ \cline{2-6} 
\multicolumn{1}{|c|}{}                                 & \multicolumn{1}{c|}{rDCNet}  & \multicolumn{1}{c|}{1.75×10$^{-7}$} & \multicolumn{1}{c|}{1.35×10$^{-2}$} & \multicolumn{1}{c|}{3.59×10$^{-5}$} & 1.46×10$^{-4}$ \\ \cline{2-6} 
\multicolumn{1}{|c|}{}                                 & \multicolumn{1}{c|}{rEEGNet} & \multicolumn{1}{c|}{2.31×10$^{-7}$} & \multicolumn{1}{c|}{4.56×10$^{-3}$} & \multicolumn{1}{c|}{2.25×10$^{-2}$} & 4.56×10$^{-3}$ \\ \hline
\end{tabular}
\end{tabular}
}
\label{tab:sensor_source_p_values}
\vspace{0.15cm}
\scriptsize{\\
rDCNet - rDeepConvNet, rSCNet - rShallowConvNet}
\end{table*}
\begin{table*}
\centering
\caption{$p$ values of one-tailed $t$-test between trajectory decoding models with sensor-domain and source-domain features for Intra-subject setting in the x, y, and z directions.}
\begin{tabular}{ll}
\begin{tabular}{ccccc|}
\hline \hline
\multicolumn{5}{c}{Sensor-domain}\\
\hline \hline
\multicolumn{5}{c}{x-direction}\\
\hline \hline
\multicolumn{1}{|c|}{}       & \multicolumn{1}{c|}{mLR}                         & \multicolumn{1}{c|}{rSCNet}                       & \multicolumn{1}{c|}{rDCNet}                       & rEEGNet                      \\ \hline
\multicolumn{1}{|c|}{mLR}    & \multicolumn{1}{c|}{-}                           & \multicolumn{1}{c|}{2.90×10$^{-6}$} & \multicolumn{1}{c|}{6.00×10$^{-6}$} & 2.66×10$^{-7}$ \\ \hline
\multicolumn{1}{|c|}{rSCNet}  & \multicolumn{1}{c|}{2.90×10$^{-6}$} & \multicolumn{1}{c|}{-}                           & \multicolumn{1}{c|}{5.91×10$^{-2}$} & 7.46×10$^{-5}$ \\ \hline
\multicolumn{1}{|c|}{rDCNet}  & \multicolumn{1}{c|}{6.00×10$^{-6}$} & \multicolumn{1}{c|}{5.91×10$^{-2}$} & \multicolumn{1}{c|}{-}                           & 1.56×10$^{-3}$ \\ \hline
\multicolumn{1}{|c|}{rEEGNet} & \multicolumn{1}{c|}{2.66×10$^{-7}$} & \multicolumn{1}{c|}{7.46×10$^{-5}$} & \multicolumn{1}{c|}{1.56×10$^{-3}$} & -                           \\ \hline \hline
\multicolumn{5}{c}{y-direction}                                                                                                                                                                                   \\ \hline \hline
\multicolumn{1}{|c|}{}       & \multicolumn{1}{c|}{mLR}                         & \multicolumn{1}{c|}{rSCNet}                       & \multicolumn{1}{c|}{rDCNet}                       & rEEGNet                      \\ \hline
\multicolumn{1}{|c|}{mLR}    & \multicolumn{1}{c|}{-}                           & \multicolumn{1}{c|}{1.55×10$^{-6}$} & \multicolumn{1}{c|}{7.72×10$^{-7}$} & 2.98×10$^{-7}$ \\ \hline
\multicolumn{1}{|c|}{rSCNet}  & \multicolumn{1}{c|}{1.55×10$^{-6}$} & \multicolumn{1}{c|}{-}                           & \multicolumn{1}{c|}{5.43×10$^{-4}$} & 2.04×10$^{-5}$ \\ \hline
\multicolumn{1}{|c|}{rDCNet}  & \multicolumn{1}{c|}{7.72×10$^{-7}$} & \multicolumn{1}{c|}{5.43×10$^{-4}$} & \multicolumn{1}{c|}{-}                           & 6.94×10$^{-5}$ \\ \hline
\multicolumn{1}{|c|}{rEEGNet} & \multicolumn{1}{c|}{2.98×10$^{-7}$} & \multicolumn{1}{c|}{2.04×10$^{-5}$} & \multicolumn{1}{c|}{6.94×10$^{-5}$} & -                           \\ \hline \hline
\multicolumn{5}{c}{z-direction}                                                                                                                                                                                   \\ \hline \hline
\multicolumn{1}{|c|}{}       & \multicolumn{1}{c|}{mLR}                         & \multicolumn{1}{c|}{rSCNet}                       & \multicolumn{1}{c|}{rDCNet}                       & rEEGNet                      \\ \hline
\multicolumn{1}{|c|}{mLR}    & \multicolumn{1}{c|}{-}                           & \multicolumn{1}{c|}{1.88×10$^{-7}$} & \multicolumn{1}{c|}{2.55×10$^{-7}$} & 3.13×10$^{-9}$ \\ \hline
\multicolumn{1}{|c|}{rSCNet}  & \multicolumn{1}{c|}{1.88×10$^{-7}$} & \multicolumn{1}{c|}{-}                           & \multicolumn{1}{c|}{7.89×10$^{-2}$} & 6.33×10$^{-4}$ \\ \hline
\multicolumn{1}{|c|}{rDCNet}  & \multicolumn{1}{c|}{2.55×10$^{-7}$} & \multicolumn{1}{c|}{7.89×10$^{-2}$} & \multicolumn{1}{c|}{-}                           & 3.55×10$^{-4}$ \\ \hline
\multicolumn{1}{|c|}{rEEGNet} & \multicolumn{1}{c|}{3.13×10$^{-9}$} & \multicolumn{1}{c|}{6.33×10$^{-4}$} & \multicolumn{1}{c|}{3.55×10$^{-4}$} & -                           \\ \hline
\end{tabular}
&
\begin{tabular}{ccccc|}
\hline \hline
\multicolumn{5}{c}{Source-domain}\\
\hline \hline
\multicolumn{5}{c}{x-direction}\\
\hline \hline
\multicolumn{1}{|c|}{}        & \multicolumn{1}{c|}{mLR}      & \multicolumn{1}{c|}{rSCNet}   & \multicolumn{1}{c|}{rDCNet}   & rEEGNet  \\ \hline
\multicolumn{1}{|c|}{mLR}     & \multicolumn{1}{c|}{-}        & \multicolumn{1}{c|}{8.69×10$^{-7}$} & \multicolumn{1}{c|}{3.38×10$^{-7}$} & 9.79×10$^{-8}$ \\ \hline
\multicolumn{1}{|c|}{rSCNet}  & \multicolumn{1}{c|}{8.69×10$^{-7}$} & \multicolumn{1}{c|}{-}        & \multicolumn{1}{c|}{6.43×10$^{-4}$} & 3.99×10$^{-5}$ \\ \hline
\multicolumn{1}{|c|}{rDCNet}  & \multicolumn{1}{c|}{3.38×10$^{-7}$} & \multicolumn{1}{c|}{6.43×10$^{-4}$} & \multicolumn{1}{c|}{-}        & 5.94×10$^{-5}$ \\ \hline
\multicolumn{1}{|c|}{rEEGNet} & \multicolumn{1}{c|}{9.79×10$^{-8}$} & \multicolumn{1}{c|}{3.99×10$^{-5}$} & \multicolumn{1}{c|}{5.94×10$^{-5}$} & -        \\ \hline \hline
\multicolumn{5}{c}{y-direction}                                                                                                                                                                                   \\ \hline \hline
\multicolumn{1}{|c|}{}        & \multicolumn{1}{c|}{mLR}      & \multicolumn{1}{c|}{rSCNet}   & \multicolumn{1}{c|}{rDCNet}   & rEEGNet  \\ \hline
\multicolumn{1}{|c|}{mLR}     & \multicolumn{1}{c|}{-}        & \multicolumn{1}{c|}{1.73×10$^{-7}$} & \multicolumn{1}{c|}{6.20×10$^{-8}$} & 7.57×10$^{-8}$ \\ \hline
\multicolumn{1}{|c|}{rSCNet}  & \multicolumn{1}{c|}{1.73×10$^{-7}$} & \multicolumn{1}{c|}{-}        & \multicolumn{1}{c|}{1.04×10$^{-4}$} & 8.73×10$^{-6}$ \\ \hline
\multicolumn{1}{|c|}{rDCNet}  & \multicolumn{1}{c|}{6.20×10$^{-8}$} & \multicolumn{1}{c|}{1.04×10$^{-4}$} & \multicolumn{1}{c|}{-}        & 3.03×10$^{-5}$ \\ \hline
\multicolumn{1}{|c|}{rEEGNet} & \multicolumn{1}{c|}{7.57×10$^{-8}$} & \multicolumn{1}{c|}{8.73×10$^{-6}$} & \multicolumn{1}{c|}{3.03×10$^{-5}$} & -        \\ \hline \hline
\multicolumn{5}{c}{z-direction}                                                                                                                                                                                   \\ \hline \hline
\multicolumn{1}{|c|}{}        & \multicolumn{1}{c|}{mLR}      & \multicolumn{1}{c|}{rSCNet}   & \multicolumn{1}{c|}{rDCNet}   & rEEGNet  \\ \hline
\multicolumn{1}{|c|}{mLR}     & \multicolumn{1}{c|}{-}        & \multicolumn{1}{c|}{9.95×10$^{-9}$} & \multicolumn{1}{c|}{3.98×10$^{-7}$} & 8.13×10$^{-9}$ \\ \hline
\multicolumn{1}{|c|}{rSCNet}  & \multicolumn{1}{c|}{9.95×10$^{-9}$} & \multicolumn{1}{c|}{-}        & \multicolumn{1}{c|}{7.53×10$^{-3}$} & 7.97×10$^{-7}$ \\ \hline
\multicolumn{1}{|c|}{rDCNet}  & \multicolumn{1}{c|}{3.98×10$^{-7}$} & \multicolumn{1}{c|}{7.53×10$^{-3}$} & \multicolumn{1}{c|}{-}        & 2.52×10$^{-4}$ \\ \hline
\multicolumn{1}{|c|}{rEEGNet} & \multicolumn{1}{c|}{8.13×10$^{-9}$} & \multicolumn{1}{c|}{7.97×10$^{-7}$} & \multicolumn{1}{c|}{2.52×10$^{-4}$} & -        \\ \hline
\end{tabular}
\end{tabular}
\label{tab:intrasub_p_values}
\vspace{0.15cm}
\scriptsize{\\
rDCNet - rDeepConvNet, rSCNet - rShallowConvNet}
\end{table*}

\subsubsection{Sensor-domain and Source-domain Feature Analysis}
EEG sensor-domain and source-domain features are employed as inputs to the decoding models to access the suitable features for MKD during the grasp-and-lift task. In intra-subject decoding analysis, the highest correlation values are obtained by deep learning models while using EEG sensor-domain features in x and y-direction decoding. However, in z-direction, the decoding performance of the models is better with EEG source-domain features. In particular, the rEEGNet neural decoder performs best with both sensor-domain and source-domain features as input. In inter-subject decoding analysis, it is observed that the best correlation values are obtained using sensor-domain EEG features with the rEEGNet decoder. However, the rEEGNet model with source-domain features has better decoding performance in the z-direction. A detailed $t$-test is performed to compare the decoding performance between the decoding models with EEG sensor-domain and source-domain input features. The results of the $t$-test are shown in Table \ref{tab:sensor_source_p_values} in the form of $p$-values for intra-subject and inter-subject settings.

\subsubsection{Intra-subject and Inter-subject Decoding Analysis}
The objective of the intra-subject and inter-subject decoding analysis is to explore the subject-specific and subject-independent feature-learning capabilities of the decoding model for the kinematics trajectory prediction during the grasp-and-lift task. It can be noted that the decoding model, in particular rEEGNet, is able to learn subject-specific features for MKP with sensor-domain as well as source-domain EEG features as input during the grasp-and-lift task, as can be depicted from the results in Table \ref{tab:intrasub_pccresults}. The mean correlation values of $0.741$ and $0.731$ across three axes are obtained, with $100$ $ms$ EEG lag and $450$ $ms$ window size for sensor-domain and source-domain features as decoder input, respectively. We can conclude that the rEEGNet model can learn the subject-specific motor information from the EEG sensor-domain and source-domain features for MKP.

In inter-subject decoding analysis, subject-independent features are learned by the decoding models, and the performance evaluation is done on the subject data that is excluded from the training phase. The rEEGNet model with sensor-domain input features has the best decoding performance in the x and y directions. However, in the z-direction, source-domain input features based rEEGNet have better decoding performance, as shown in Table \ref{tab:sensor_source_p_values}.



\section{Conclusion}\label{sec:conclusion}
This study explores the EEG source imaging-based kinematics decoding using a deep learning-based decoding model. In particular, the frontoparietal regions are selected for the MKP during the grasp-and-lift task. The motor-related information encoded in the pre-movement brain activation is utilized for hand trajectory decoding using the time-lagged EEG features in sensor and source domains. The MKP analysis is performed using the convolutional neural network-based decoding models. Further, inter-subject decoding analysis has been performed to evaluate the subject-independent feature-learning capabilities of the decoding models. The proposed rEEGNet yielded the overall best kinematics decoding performance using the EEG sensor-domain features. The Pearson correlation coefficient (PCC) is used as a performance metric for MKP. The decoding analysis shows the viability of continuous trajectory decoding using EEG source domain features. However, the decoding accuracy using the sensor-domain EEG features is better than the source-domain counterpart using the proposed decoding model.

\section*{Acknowledgment}
The authors would like to thank Prof. Sitikantha Roy and Prof. Shubhendu Bhasin from the Indian Institute of Technology Delhi and Dr. Suriya Prakash from All India Institute of Medical Sciences Delhi for their constructive comments during the preparation of the manuscript.

\bibliographystyle{IEEEtran}

\bibliography{ESI_v01}

\end{document}